\documentclass{article} % For LaTeX2e
\usepackage{amsmath,amsfonts,bm}

\def\eqref#1{equation~\ref{#1}}
\def\1{\bm{1}}

\DeclareMathAlphabet{\mathsfit}{\encodingdefault}{\sfdefault}{m}{sl}
\SetMathAlphabet{\mathsfit}{bold}{\encodingdefault}{\sfdefault}{bx}{n}

\newcommand{\E}{\mathbb{E}}

\usepackage{hyperref}
\usepackage{url}
\usepackage{amsmath,amssymb,amsthm,natbib}
\usepackage{booktabs}
\usepackage{graphicx}
\usepackage{multirow}
\usepackage{xcolor}
\usepackage{float}
\usepackage{tikz}
\usetikzlibrary{positioning,arrows.meta,fit}
\usepackage{algorithm}
\usepackage{algpseudocode}
\usepackage[capitalize,noabbrev]{cleveref}

\newtheorem{definition}{Definition}
\newtheorem{lemma}{Lemma}

\newcommand{\EU}{\mathrm{EU}}
\newcommand{\EUred}{\EU^{\mathrm{red}}_n}
\newcommand{\EUspr}{\EU^{\mathrm{spr}}_n}
\newcommand{\pL}{p^{L}}
\newcommand{\pU}{p^{U}}
\newcommand{\Hup}{\overline{H}}
\newcommand{\Hlo}{\underline{H}}
\newcommand{\drho}{\Delta\rho}

\title{Credal Wrapper for Uncertainty Modelling\\in Sparse Geological Data}

\author{David Jeevankumar$^1$, Maryam Sultana$^2$, Alexander D. Rast$^2$, Fabio Cuzzolin$^2$ \\
$^1$ University of Oxford \\
$^2$ AIDAS Institute, Oxford Brookes University}

\begin{document}

\maketitle

\begin{abstract}
Machine learning models for mineral exploration are trained on a few hundred sparse, expensive samples and used to decide where to drill next, so their epistemic uncertainty (EU), the part that more data could remove, matters as much as their predictions. Yet EU cannot be measured on real data, because the counterfactual that defines it is never observed, so estimators are graded by proxies such as out-of-distribution detection and calibration. We contribute a general simulation protocol that makes EU directly computable for a given model and can grade any EU estimator. It separates two estimands the literature conflates, reducible error and estimator spread, builds an oracle reference for each in a synthetic prospectivity field sampled by clustered drilling surveys, and compares uncertainty measures of an identical ensemble within each simulated world. We use it to ask when credal (imprecise-probability) measures are more faithful than Bayesian model averaging (BMA) in the sparse, low-class regime of geological classification. Three findings follow. First, the credal interval width recovers the epistemic ordering better than BMA when members are few ($\drho \approx +0.06$ to $+0.07$ in rank correlation for $M \le 5$, $P < 10^{-4}$), confirming the small-ensemble advantage reported for credal wrappers; the advantage vanishes by $M \approx 10$ and reverses at $M = 40$. Second, the choice of credal functional matters more than the choice between credal and BMA: in binary (deposit versus barren) classification the entropy interval carries no epistemic signal ($\rho \approx 0.00$) while the interval width of the same credal set does. Third, the credal advantage does not widen as training data are thinned, so it compensates for few posterior samples rather than few data. On real drill-hole geochemistry from South Australia, a resampling oracle reproduces the first finding. The protocol applies to any EU estimator whose data-generating process can be simulated.
\end{abstract}

\section{Introduction}
\label{sec:intro}

\emph{Epistemic uncertainty} (EU) \cite{Swiler09epistemicuncertainty,oren2022mcts,li_2025_esi,walters2023investor,bickfordsmith_2025_rethinking}, the part of predictive uncertainty that more data could remove (also called \emph{reducible} uncertainty), drives active learning, selective prediction and sequential experimental design \citep{kendall2017uncertainties,hullermeier2021aleatoric,settles2009active}. It matters most where data are scarce and decisions are expensive. Mineral exploration is a clear example: models trained on a few hundred irregularly placed drill samples are used to choose where to drill next; the new results update the model, which then chooses the following hole \citep{mern2024intelligent,rodriguez2018classification}. The value of this sequential cycle depends on whether the model's EU points to locations where new data would actually teach it something. In such settings ensembles are also small, because every ensemble member or posterior sample is an expensive fit.

Most practical EU estimators read uncertainty from a finite set of member predictions, whether deep-ensemble members, posterior samples of a Bayesian neural network, or dropout masks \citep{lakshminarayanan2017simple,gal2016dropout,jospin2022hands}; we call any such set an \emph{ensemble}. The standard approach is Bayesian model averaging (BMA), with epistemic uncertainty measured by the mutual-information decomposition \citep{houlsby2011bald}. 
Within an imprecise-probability framework \citep{cuzzolin2020geometry}, 
\emph{credal wrappers} \citep{wang2025credal}, instead, retain the range of member predictions as a \emph{credal set}, a convex set of distributions \cite{levi1980enterprise,caprio2025credal}. Credal-set width has been used directly as an EU measure \citep{manchingal2025epistemic}. Credal wrappers were shown to give more reliable uncertainty than BMA when only a few samples are available. We use the post-hoc credal wrapper because it applies to any existing ensemble without retraining; our conclusions apply to any method that outputs per-class probability intervals.

Any claim of this kind faces a basic measurement problem. EU is defined by a counterfactual: how would the prediction change if the model had been trained on different data of the same kind? On real data that counterfactual is never observed, so EU has no reference value and is evaluated through proxies such as out-of-distribution detection and calibration. These proxies are informative but do not isolate the epistemic estimate, and recent work argues that evaluation should target a quantity such as the reducible error of the deployed predictor \citep{paplham2026evaluating,jimenez2026position}. A second, quieter problem is that ``epistemic uncertainty'' names at least two different quantities: the excess error that more data would remove \citep{hullermeier2021aleatoric,lahlou2023deup}, and the spread of the estimator under resampling \citep{houlsby2011bald,depeweg2018decomposition}. They usually agree and diverge exactly when all members are confidently wrong together.

For example, if a survey never samples a barren granite next to a drilled ore body, every ensemble member extrapolates the mineralisation into the granite in the same way: the members agree, so the spread is near zero, yet all are wrong, so the reducible error is large. An ensemble trained on scans from one hospital and applied to another scanner fails in the same way. Disagreement-based estimators, BMA and credal alike, cannot see such cases.

We address both problems with an evaluation protocol in which the data-generating process is known, so each EU estimand has a computable oracle reference for a given model (\cref{fig:protocol}). The protocol is general: it can grade any estimator that assigns an EU score to each input, whether BMA, MC-Dropout, an evidential model or a credal set, provided the data-generating process can be simulated. We apply it to a question on which the evidence so far has come only from proxies: under which conditions, and through which functional, does a credal uncertainty measure of a finite ensemble recover EU more faithfully than BMA? The paper is a methods paper with a geoscience case study. Geology sets the regime we study (few ensemble members, few classes, small and spatially clustered datasets) and supplies the real-data test.

\paragraph{Novelty.} Credal and BMA uncertainty measures have so far been compared through proxies such as out-of-distribution detection, calibration and accuracy \citep{wang2025credal,shaker2021ensemble}. Oracles from a known data-generating process have measured the reducible error of trained networks \citep{khorasani2026beyond} and studied EU estimators qualitatively \citep{jimenez2026position}. To our knowledge, no prior work (i) grades uncertainty measures computed from the same ensemble against a computed EU target, location by location and with paired statistics, (ii) builds oracles for both reducible error and estimator spread on the same worlds and measures how far they agree, or (iii) compares credal functionals in the low-class regime typical of scientific classification.

\paragraph{Contributions.}
\begin{enumerate}
\item \textbf{An oracle-referenced evaluation protocol for EU} (\cref{sec:protocol}). We distinguish EU as the reducible error of a model from EU as the spread of its predictions when it is retrained on resampled data, and build an oracle for each by retraining the model many times in a simulated world where the true class probabilities are known. Measures are scored by how well they rank locations in the oracle's order, compared within each world. The protocol applies to any EU estimator.
\item \textbf{Credal uncertainty is more faithful than BMA when the ensemble is small, and only then} (\cref{sec:ensemble}). Summarising the credal set by its interval width (for each class, the gap between the largest and smallest probability any member assigns), two to five members rank locations more like the oracle than BMA does (by about 0.06 to 0.07 in rank correlation, significantly); for our main member model the advantage disappears at roughly ten members and reverses at forty, and the small-ensemble lead reproduces on real geochemical data.
\item \textbf{How the credal set is summarised matters more than whether one is used} (\cref{sec:functional}). The entropy interval used by \citet{wang2025credal}, the difference between the highest and lowest entropy of any distribution in the set, shows no agreement with the oracle in two-class problems, while the width of the same set does; it improves as the number of classes grows. It depends on where the set lies as well as how wide it is, so it partly tracks the confidence of the average prediction, which reflects label noise, rather than member disagreement, which more data would resolve.
\item \textbf{The advantage depends on the number of members, not the amount of data} (\cref{sec:sparsity}). Within the range tested, shrinking the training set from 500 to 100 samples leaves the credal advantage unchanged.
\end{enumerate}

\paragraph{Outline.} \Cref{sec:related} reviews related work; \cref{sec:protocol} defines the estimands, oracles and grading; \cref{sec:readouts} the uncertainty measures; \cref{sec:setup,sec:results} the setup and results, including real geochemistry; \cref{sec:discussion} guidance and limitations.

\section{Related work}
\label{sec:related}

\paragraph{Random-set and credal prediction.}
Within imprecise probability applied to machine learning, two prominent approached are based on random sets and credal sets, respectively.

\emph{Random-set models}, within the approach termed `epistemic AI' (\cite{cuzzolin2024epistemic,manchingal2024epistemic,manchingal2025epistemic}), allow probability to be assigned to sets of possible
outcomes without specifying how it is divided among their members
\citep{shafer2020mathematical,cuzzolin2020geometry}.
Random-Set Neural Networks (RS-NN) implement this representation for
classification, including text classification
\citep{manchingal2025rsnn}.
Random-Set Large Language Models (RS-LLM) extend it to language generation,
predicting over sets of possible next tokens
\citep{mubashar2025rsllm}.
Other approaches obtain lower and upper probabilities without predicting
probabilities for answer sets, for instance those using credal sets, i.e., convex sets of distributions \cite{levi1980enterprise,caprio2025credal}. 

Within \emph{credal set} representations, Credal Deep Ensembles combine networks
trained to predict probability bounds \citep{wang2024credalensembles,wang2026learning,wang2026credal},
while the Credal Wrapper constructs bounds from the predictions of
Bayesian neural networks or deep ensembles
\citep{wang2025credalwrapper}, using the intersection probability (\cite{cuzzolin2009credal,cuzzolin2007properties}) as pointwise estimator. Credal INNs \cite{wang2025creinns} combine the expressive power of interval neural networks with credal representations.
For language models, \citet{yang2026imprecise} investigate prompting
models to report lower and upper probabilities for possible answers,
as well as constructing intervals from multiple reported predictions.
The credal approach has been recently extended to large language models \cite{manchingal2026credal}, generative adversarial networks (\cite{mubashar2026epistemic}) and to a wrapper applicable to Bayesian NN (\cite{sultana2025epistemic}).
Our classifiers follow the random-set approach of RS-NN and RS-LLM (\cite{mubashar2025rsllm}),
but predict over four answer options rather than vocabulary tokens.
Second-order probability representation has been applied in the past to pose estimation (\cite{gong2017belief}), tracking (\cite{cuzzolin1999evidential}), logistic regression (\cite{cuzzolin2018belief}), clustering (\cite{zhu2026tdcc}), feature fusion (\cite{pang2025guest}), max-entropy classification (\cite{cuzzolin2018generalised}), statistical learning theory (\cite{cuzzolin2024generalising,caprio2024credal}) and a variety of other tasks (\cite{liu2019evidence,faza2026direct,manchingal2025uncertainty}).
A recent review can be found in \cite{wang2025review}.
We investigate how input tokens influence the resulting credal width,
rather than proposing a new uncertainty representation.

\paragraph{Ensemble-based EU and credal representations.}
Credal approaches thus keep a set of distributions rather than averaging ensemble members. \citet{shaker2021ensemble} compare Bayesian and credal ensemble inference; \citet{wang2025credal} introduce the post-hoc credal wrapper studied here; related work learns interval-valued outputs \citep{wang2024credal,wang2025creinns}, distils credal ensembles \citep{wang2026distillation} or uses credal priors \citep{caprio2024credal} (survey: \citealp{wang2026review}). \emph{Evidential methods} \cite{gao2025comprehensive} offer a different second-order representation based on Dirichlet distributions \citep{sensoy2018evidential}. Random-set neural networks \citep{manchingal2023random} predict a belief function \citep{shafer2020mathematical,cuzzolin2014belief,cuzzolin2018visions,cuzzolin2020geometry,cuzzolin2026statistical} over sets of classes, which induces a credal set whose width can serve as an EU measure \citep{manchingal2025epistemic}.
A geometric approach to belief functions and other uncertainty measures can be found in \cite{cuzzolin2004geometry,cuzzolin2008geometric,cuzzolin2008credal,cuzzolin2020geometry-dempster}; their combinatorial analysis in \cite{cuzzolin2010three}, their algebraic properties in \cite{cuzzolin01bcc,cuzzolin2001lattice,zhou2017total}.
Their outputs could be graded with the same protocol. \emph{None of our results depend on the specific wrapper}: the protocol grades any per-input EU score, and the functional comparison applies to any method that outputs per-class probability intervals, including credal deep ensembles, interval networks and random-set networks.

\paragraph{Functionals for credal sets.} A credal set can be reduced to a scalar by a functional: non-specificity \citep{abellan2000nonspecificity}, upper and lower entropy \citep{abellan2003maximum}, volume, or distance-based measures \citep{sale2024second}. \citet{sale2023volume} show that volume is a sensible EU measure for binary classification but less so for many classes; as we will see in the following, our entropy-interval result runs the other way.

\paragraph{Evaluating EU.}
Analytical work has shown that common EU measures can violate desirable properties \citep{wimmer2023quantifying,schweighofer2025information} and that EU learned through loss minimisation need not be faithful \citep{bengs2022pitfalls}. \citet{paplham2026evaluating} argue that the right target is the expected regret of the deployed predictor. \citet{khorasani2026beyond} use class-conditional normalizing flows as exact-posterior oracles on images to measure the aleatoric floor and the reducible error of trained networks, and \citet{jimenez2026position} use simulated data-generating processes to study EU estimators qualitatively. We use such oracles instead to grade EU estimators.

\paragraph{EU in the geosciences.}
Mineral prospectivity mapping predicts where deposits are likely from sparse, expensive observations: drill holes, geochemical assays of soil and rock, and geophysical surveys. Datasets typically hold a few hundred to a few thousand samples, clustered around past exploration targets, and labels are few: deposit versus barren, or a handful of rock types. Machine learning, including Bayesian deep learning, is widely used for these tasks \citep{rodriguez2018classification,yuan2026systematic,huang2022geochemical,liu2025quantifying}. Because each hole is costly, exploration is increasingly framed as sequential decision-making under epistemic uncertainty, where the model's EU decides the next hole \citep{mern2024intelligent}. In this literature EU quality is assessed only with proxies, and spatial autocorrelation is often handled by random splits, which inflate performance \citep{roberts2017cross}; we address the first with the oracle protocol and measure the second on real data.

\section{Oracle-referenced evaluation of epistemic uncertainty}
\label{sec:protocol}

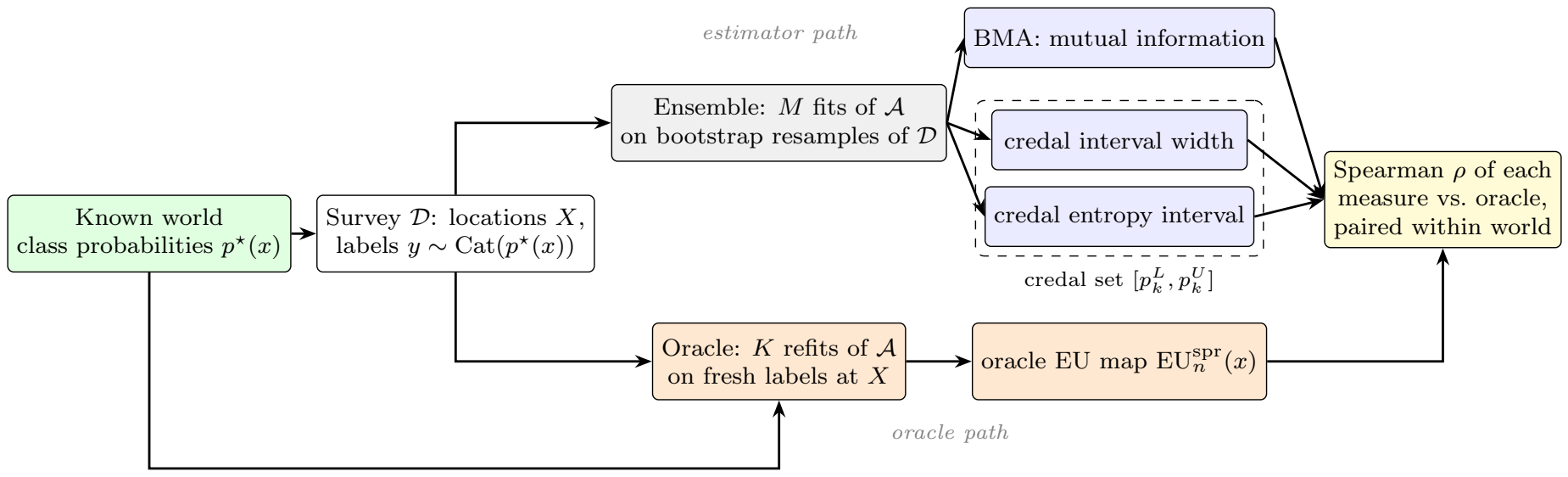
\begin{figure}[t]
\centering
\resizebox{\linewidth}{!}{%
\begin{tikzpicture}[
  box/.style={draw, rounded corners=2pt, align=center, font=\footnotesize, minimum height=9mm, minimum width=26mm, inner sep=3pt},
  est/.style={box, fill=gray!12},
  orc/.style={box, fill=orange!18},
  msr/.style={box, fill=blue!8, minimum width=30mm, minimum height=7mm},
  arr/.style={-{Stealth[length=2mm]}, thick}]
\node[box, fill=green!12] (world) at (0,0) {Known world\\class probabilities $p^\star(x)$};
\node[box] (data) at (3.6,0) {Survey $\mathcal{D}$: locations $X$,\\labels $y\sim\mathrm{Cat}(p^\star(x))$};
\node[est] (ens) at (7.4,1.3) {Ensemble: $M$ fits of $\mathcal{A}$\\on bootstrap resamples of $\mathcal{D}$};
\node[orc] (orc) at (7.4,-1.5) {Oracle: $K$ refits of $\mathcal{A}$\\on fresh labels at $X$};
\node[msr] (bma) at (11.4,2.3) {BMA: mutual information};
\node[msr] (wid) at (11.4,1.1) {credal interval width};
\node[msr] (ent) at (11.4,0.2) {credal entropy interval};
\node[draw, dashed, rounded corners=3pt, fit=(wid)(ent), inner sep=3pt, label={[font=\scriptsize]below:credal set $[p^L_k, p^U_k]$}] (cs) {};
\node[orc, minimum width=30mm] (eur) at (11.4,-1.5) {oracle EU map $\mathrm{EU}^{\mathrm{spr}}_n(x)$};
\node[box, fill=yellow!20, minimum width=24mm] (rho) at (15.2,0.4) {Spearman $\rho$ of each\\measure vs.\ oracle,\\paired within world};
\draw[arr] (world) -- (data);
\draw[arr] (data.north) |- (ens.west);
\draw[arr] (data.south) |- (orc.west);
\draw[arr] (world.south) |- ([yshift=-8mm]orc.south) -- (orc.south);
\draw[arr] (ens.east) -- (bma.west);
\draw[arr] (ens.east) -- (wid.west);
\draw[arr] (ens.east) -- (ent.west);
\draw[arr] (orc) -- (eur);
\draw[arr] (bma.east) -- (rho.west);
\draw[arr] (wid.east) -- (rho.west);
\draw[arr] (ent.east) -- (rho.west);
\draw[arr] (eur.east) -| (rho.south);
\node[font=\scriptsize\itshape, text=gray] at (7.4,2.35) {estimator path};
\node[font=\scriptsize\itshape, text=gray] at (9.4,-2.35) {oracle path};
\end{tikzpicture}}
\caption{The evaluation protocol. On the estimator path, an ensemble is fitted on bootstrap resamples of a survey drawn from a known world, and its predictions are summarised by each uncertainty measure. On the oracle path, the same algorithm $\mathcal{A}$ is refitted on fresh labels at the same locations, giving the target EU map. Each measure is scored by its Spearman correlation with the oracle map, compared within each world.}
\label{fig:protocol}
\end{figure}

This section defines what we measure and how we compute it. \Cref{fig:protocol} gives an overview, and \cref{alg:protocol} in \cref{app:alg} gives the full procedure as pseudocode.

\subsection{Setting and notation}
\label{sec:notation}

We consider classification with $C$ classes. A \emph{dataset} $\mathcal{D} = \{(x_i, y_i)\}_{i=1}^n$ consists of inputs $x_i \in \mathcal{X}$ and labels $y_i \in \{1,\dots,C\}$. In the geological setting an input is a location, a label is the rock type or deposit status found by drilling there, and a dataset is a drilling \emph{survey}; we use the two words interchangeably. We call the set of locations $X = \{x_1,\dots,x_n\}$ the \emph{design}, as distinct from the labels observed there. Each location has a true class-probability vector $p^\star(x) \in \Delta^{C-1}$, where $\Delta^{C-1} = \{p \in [0,1]^C : \sum_k p_k = 1\}$ is the probability simplex, and a drilled location returns a random label $y \sim \mathrm{Cat}(p^\star(x))$. We write $\mathcal{P}$ for the process that generates datasets in this way.

A \emph{learning algorithm} $\mathcal{A}$ is a training procedure (model class, loss and optimiser) that maps a dataset to a fitted \emph{model} $f = \mathcal{A}(\mathcal{D})$, which returns a predicted probability vector $f(x) \in \Delta^{C-1}$ at any location. A \emph{refit} runs the same algorithm again on a different dataset. An \emph{ensemble} is a set of $M$ models $f_1,\dots,f_M$ fitted by $\mathcal{A}$ on bootstrap resamples of $\mathcal{D}$; its \emph{members} predict $p_m(x) = f_m(x)$. An \emph{uncertainty measure} \cite{cuzzolin2020geometry,cuzzolin2021uncertainty,cuzzolin2025uncertainty} maps the member predictions $p_1(x),\dots,p_M(x)$ to a single EU score. $H(p) = -\sum_k p_k \log_2 p_k$ is the Shannon entropy in bits.

\subsection{Two definitions of epistemic uncertainty}

Neither definition below is new. What we add is a way to compute both, so that estimators can be graded against them (\cref{sec:oracles}).

\begin{definition}[EU as reducible error]\label{def:red}
$\EUred(x) = \E_{\mathcal{D}\sim\mathcal{P}^n}\big[\|f_n(x) - p^\star(x)\|^2\big]$, where $f_n = \mathcal{A}(\mathcal{D})$.
\end{definition}

The expected squared (Brier) loss of a prediction $f(x)$ against a random label splits into two parts: $\E_{y}\|f(x) - e_y\|^2 = \|f(x) - p^\star(x)\|^2 + \big(1 - \|p^\star(x)\|^2\big)$, where $e_y$ is the one-hot vector of $y$. The second term is the \emph{aleatoric floor}: the loss that even the Bayes-optimal predictor $p^\star$ incurs because labels are random, which no amount of data can remove \citep{khorasani2026beyond}. Under log loss the floor is $H(p^\star(x))$. The first term is the part due to training on finite data; averaging it over datasets gives \cref{def:red}, the excess risk used by \citet{lahlou2023deup} and \citet{paplham2026evaluating}, which vanishes as $n$ grows for a consistent algorithm. We use squared error rather than log loss for \cref{def:red} because its excess over the floor has the closed form $\|f(x) - p^\star(x)\|^2$ and splits into bias and variance; the two oracles are compared by rank, so the difference in scale from \cref{def:spr} does not matter. For example, next to dense drilling any model predicts close to $p^\star(x)$, so reducible error is small even where the label is a coin flip; far from any hole, predictions depend heavily on which survey was drawn, and reducible error is large.

\begin{definition}[EU as estimator spread]\label{def:spr}
Let $q_1(x),\dots,q_K(x) \in \Delta^{C-1}$ be the predictions at $x$ of $K$ refits of $\mathcal{A}$ on datasets with the same locations and fresh labels drawn from $p^\star$, and $\bar q(x) = \frac1K \sum_k q_k(x)$ their average. Then $\EUspr(x) = H(\bar q(x)) - \frac1K \sum_{k=1}^K H(q_k(x))$.
\end{definition}

This is the mutual-information decomposition used by BALD \citep{houlsby2011bald,depeweg2018decomposition}, applied to refits instead of posterior samples. The average $\bar q$ is the prediction obtained by pooling the refits. \Cref{def:spr} is the entropy of that pooled prediction minus the average entropy of the individual refits, so it is zero when all refits agree and grows as they disagree. For example, two refits predicting $(0.9, 0.1)$ and $(0.1, 0.9)$ pool to $(0.5, 0.5)$, and $\EUspr = 1 - 0.47 = 0.53$ bits. Two refits that both predict $(0.5, 0.5)$ give $\EUspr = 0$: each is uncertain, but that uncertainty is aleatoric and more data would not remove it.

\paragraph{When the two disagree.} Reducible error splits into squared bias and variance \citep{geman1992neural}: $\EUred(x) = \|\E_{\mathcal{D}} f_n(x) - p^\star(x)\|^2 + \E_{\mathcal{D}}\|f_n(x) - \E_{\mathcal{D}} f_n(x)\|^2$. Estimator spread reflects only the variance part. The two therefore agree wherever bias is small, and diverge under \emph{shared bias}: every refit makes the same error, so spread is near zero while reducible error is large. This is why variance-based estimators can miss part of epistemic uncertainty \citep{lahlou2023deup,jimenez2026position}. A model class too smooth to represent a narrow vein, for instance, smooths it out identically in every refit. \Cref{fig:world} (bottom left) shows where shared bias occurs in one of our worlds.

\paragraph{Two properties of the targets.} First, both are relative to a learning algorithm and a sample size \citep{paplham2026evaluating}: EU is a property of a particular model trained on finite data, not of the world, and two algorithms trained on the same survey can be uncertain in different places. Every oracle is therefore built by refitting the same algorithm whose ensemble is being evaluated. Second, both refer to repeated draws of data from $\mathcal{P}$, not to a posterior over parameters; were EU defined as posterior spread, BMA would match the target by construction, whereas this definition favours neither method.

\subsection{A controlled generative world}
\label{sec:world}

\begin{figure}[t]
\centering
\includegraphics[width=1\linewidth]{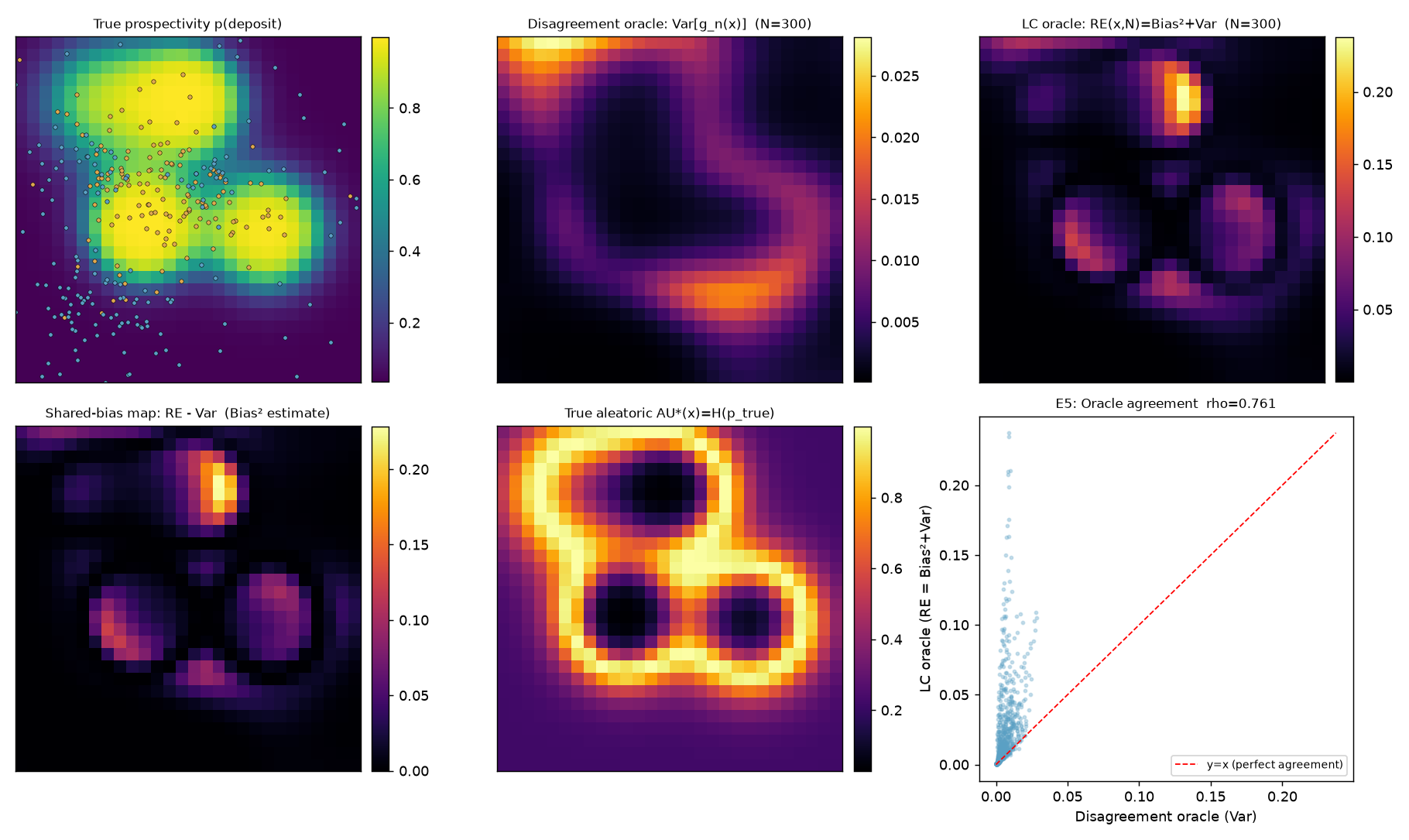}
\caption{One synthetic world ($C=2$, $n=300$). Top row: true probability of a deposit $p^\star(x)$ with the drilled locations (left); disagreement oracle, here in its variance form (centre); learning-curve oracle, reducible error $=$ bias$^2 +$ variance (right). Bottom row: their difference, which marks shared-bias locations (left); aleatoric uncertainty $H(p^\star(x))$, high along the boundaries between bodies (centre); the two oracles plotted against each other over the query grid, Spearman $\rho = 0.761$ (right).}
\label{fig:world}
\end{figure}

To compute either definition we need a world in which $p^\star$ is known and new surveys can be drawn at will. We use our own construction, chosen to be simple and cheap to sample rather than geologically realistic. Locations lie in the unit square $\mathcal{X} = [0,1]^2$. Each class $c$ has a latent score $z_c(x) = \sum_{b=1}^{B} a_{cb}\exp\!\big(-\|x-\mu_{cb}\|^2 / 2\sigma_{cb}^2\big)$, a sum of $B$ Gaussian bumps with random centres $\mu_{cb}$, widths $\sigma_{cb}$ and amplitudes $a_{cb}$. Each bump can be read as a body of one rock type. The class probabilities are $p^\star(x) = \mathrm{softmax}(\beta z(x))$, where the inverse temperature $\beta > 0$ controls how sharp the boundaries between bodies are: large $\beta$ gives near-certain labels and a low aleatoric floor, small $\beta$ gives blurred boundaries and noisy labels. Surveys are drawn around a few random hubs, mimicking the clustered coverage of real drilling campaigns, and each drilled location returns a random label from $p^\star$. \Cref{fig:world} shows one world; parameters are in \cref{app:settings}.

\subsection{Two oracles}
\label{sec:oracles}

An \emph{oracle} here is a Monte Carlo estimate of one of the two definitions, made possible because the world is known. Unlike earlier oracles \citep{khorasani2026beyond,jimenez2026position}, we build one for each definition on the same worlds and use them to grade uncertainty measures.

\paragraph{Disagreement oracle (\cref{def:spr}).} Keep the design $X$ fixed, draw fresh labels $y^{(k)}_i \sim \mathrm{Cat}(p^\star(x_i))$ for $k = 1,\dots,K$, refit $q_k = \mathcal{A}(X, y^{(k)})$ on each, and evaluate \cref{def:spr} at every point of a query grid. In words: if we drilled the same holes again and the outcomes came out differently, how much would the model change? A bootstrap ensemble estimates exactly this by resampling one dataset; the oracle computes it with genuinely new data, so a measure that disagrees with it is failing at its own task.

\paragraph{Learning-curve oracle (\cref{def:red}).} Draw $R$ independent surveys of size $n$ (new locations and new labels), fit $\mathcal{A}$ on each, and average $\|f(x) - p^\star(x)\|^2$ over the fits at every grid point. Repeating this at $n_0, 2n_0, 4n_0, \dots$ traces a learning curve, which lets us check that the estimate really decays with $n$.

The learning-curve oracle captures bias as well as variance, while the disagreement oracle captures only the spread. Where reducible error is high but refits agree lies shared bias, which no disagreement-based measure can detect; \cref{sec:robust} reports how closely the two maps agree.

\subsection{Grading an uncertainty measure}
\label{sec:grading}

\paragraph{What matters is the ranking.} In sequential acquisition, a form of active learning \citep{settles2009active}, the model is retrained after each drilling round, scores every candidate location, and the top-scoring one is drilled next \citep{mern2024intelligent}; only the order of the scores matters. We therefore score a measure by the Spearman rank correlation $\rho$ \citep{spearman1904proof} between its EU map and the oracle map over the $G$ locations of the query grid. $\rho$ is the ordinary correlation between the two sets of ranks: $1$ means the same ordering, $0$ no relation, and $-1$ the reverse ordering. We call $\rho$ the \emph{fidelity} of the measure.

\paragraph{Comparing measures within each world.} Results are averaged over $S$ independently generated worlds, indexed by $s$. Within world $s$, all measures are computed from the same ensemble and compared with the same oracle, so the fidelities $\rho^{(s)}_A$ and $\rho^{(s)}_B$ of two measures $A$ and $B$ (for example credal width and BMA) are \emph{paired}: a world that is hard for one measure tends to be hard for the other. We therefore analyse the difference within each world, $\drho^{(s)} = \rho^{(s)}_A - \rho^{(s)}_B$, which cancels the variation in difficulty between worlds, much as a paired $t$-test does. We report three numbers: the mean of $\drho^{(s)}$ over worlds; a 95\% bootstrap interval, obtained by resampling the $S$ differences with replacement 20{,}000 times and taking the 2.5th and 97.5th percentiles of the resampled means \citep{efron1994introduction}; and the $P$-value of a Wilcoxon signed-rank test \citep{wilcoxon1945individual}, which asks whether the differences are centred at zero without assuming they are normally distributed. $P$ is the probability of a difference at least this large if the two measures were equally faithful (capital $P$ avoids confusion with probability vectors $p$).

\section{Uncertainty measures under test}
\label{sec:readouts}

Each member's prediction $p_m(x)$ is a probability vector over the $C$ classes, i.e.\ a point in the simplex $\Delta^{C-1}$ (\cref{sec:notation}). All uncertainty measures below take the same set $\{p_1(x),\dots,p_M(x)\}$, so any difference between them is due to the measure alone.

\paragraph{BMA.} $\EU_{\mathrm{BMA}} = H(\bar p) - \frac1M\sum_m H(p_m)$ with $\bar p = \frac1M \sum_m p_m$: the members are averaged first and the uncertainty is read from the blend, as in \cref{def:spr}.

\paragraph{Credal set.} Following \citet{wang2025credal}, per-class bounds $\pL_k = \min_m p_{m,k}$ and $\pU_k = \max_m p_{m,k}$ define the credal set $\mathcal{P} = \{p \in \Delta^{C-1} : \pL_k \le p_k \le \pU_k\ \forall k\}$, the set of all distributions consistent with the range of member predictions. Any method that outputs lower and upper class probabilities induces the same kind of set, so the two functionals below apply equally to those methods. We summarise the set in two ways:
\begin{align}
\EU_{\mathrm{width}} &= \tfrac{1}{C}\textstyle\sum_{k=1}^{C}\big(\pU_k - \pL_k\big), \label{eq:width}\\
\EU_{\mathrm{ent}} &= \Hup(\mathcal{P}) - \Hlo(\mathcal{P}), \quad \Hup = \max_{p\in\mathcal{P}} H(p),\ \ \Hlo = \min_{p\in\mathcal{P}} H(p). \label{eq:ent}
\end{align}
The width depends only on how far apart the members are, so it is a pure disagreement measure, in line with its use as an EU measure in random-set networks \citep{manchingal2025epistemic}. The entropy interval is the functional used by \citet{wang2025credal}.

\begin{lemma}[Upper entropy by water-filling]\label{prop:water}
The distribution of maximum entropy in $\mathcal{P}$ has coordinates $p_k = \mathrm{median}(\pL_k, t, \pU_k)$, where the level $t$ is the unique value for which $\sum_k p_k = 1$.
\end{lemma}
This is a standard consequence of the KKT conditions (proof in \cref{app:proofs}); algorithms for maximum entropy over probability intervals are also given by \citet{abellan2003maximum}. The level $t$ has a simple reading, shown in \cref{fig:waterfill}: maximum entropy makes the class probabilities as equal as the intervals allow. Every class takes the common value $t$ unless its interval forces it higher or lower, and $t$ is raised until the probabilities sum to one, like water poured into columns with floors $\pL_k$ and ceilings $\pU_k$. Because $\Hup$ can be computed exactly this way for any number of classes, we can evaluate the entropy interval reliably from $C=2$ up to $C=20$ in \cref{sec:functional}. $\Hlo$ is computed by a greedy construction over the vertices of $\mathcal{P}$, which is exact for $C=2$ and approximate for $C>2$.

\paragraph{Baseline.} We also report MC-Dropout \citep{gal2016dropout}: a single network with dropout kept active at prediction time, whose $M$ stochastic forward passes are treated as ensemble members and summarised with the BMA measure. It gives a cheaper ensemble whose diversity comes from dropout masks rather than from resampling the data.

\section{Experimental setup}
\label{sec:setup}

\paragraph{Member model.} The algorithm $\mathcal{A}$ fits random-Fourier-feature multinomial logistic models \citep{rahimi2007random}: $D=128$ shared random features approximating an RBF kernel, followed by $\ell_2$-regularised logistic regression. Ensemble diversity comes only from bootstrap resampling of the drilled data, as in a deep ensemble. We chose it for two reasons. The oracles need 51{,}200 refits, feasible only if a fit takes milliseconds; random-feature logistic regression approximates a Gaussian-process classifier at that cost (under 30 CPU-minutes in total, \cref{app:compute}). And the comparison depends only on member predictions, so the member model should matter little, which \cref{sec:robust} tests with neural-network members and on real data.

\paragraph{Protocol.} Each setting uses $S=40$ independently generated worlds, a $28\times28$ query grid ($G=784$) and $K=80$ oracle refits. The oracle does not depend on $M$, so it is computed once per world and reused across ensemble sizes, which makes the pairing exact. The three sweeps are: \textbf{class count} $C \in \{2,3,5,10,20\}$ at $M=5$, with $n = 60C$ so samples per class stay roughly constant (a fixed-$n$ arm is in \cref{app:e1armA}); \textbf{number of members} $M \in \{2,3,5,10,20,40\}$ at $n = 150$, $C=2$; and \textbf{sample size}, retaining 100/80/60/40/20\% of a survey of $n=500$ at $M=5$, $C=2$.

\paragraph{Hypothesis for the class-count sweep.} The width depends only on how far apart the members are. The entropy interval also depends on where the interval sits relative to the maximum-entropy point of the simplex, so two locations with identical member disagreement can receive different scores. This is undesirable: member disagreement is what more data would reduce, whereas the interval's position largely tracks the confidence of the average prediction, which is closer to aleatoric uncertainty. In binary classification entropy is a single symmetric arc in one coordinate, and this location effect should be at its strongest. With more classes, entropy depends on the whole probability vector rather than on one coordinate's distance from $1/2$, and the effect should dilute. To our knowledge this has not been noted before for the entropy interval (cf.\ \citealp{wimmer2023quantifying}).

\section{Results}
\label{sec:results}

\begin{figure}[t]
\centering
\includegraphics[width=\linewidth]{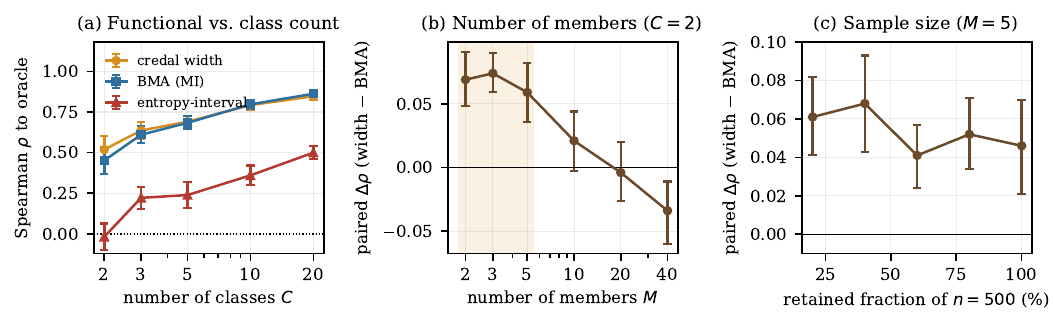}
\vspace{-1.5em}
\caption{Fidelity to the disagreement oracle (40 worlds per point). (a) Fidelity of each measure versus class count at fixed samples per class, $M=5$; bars are 95\% intervals on the mean. (b, c) Paired per-world gap between credal width and BMA with bootstrap 95\% intervals; shading marks $M \le 5$.}
\label{fig:main}
\end{figure}

\subsection{Functional choice depends on class count}
\label{sec:functional}

\Cref{fig:main}a shows three results (full numbers in \cref{tab:e1}, \cref{app:e1armA}).

\emph{The entropy interval carries no epistemic signal in binary classification.} At $C=2$ its fidelity is $-0.018$ (and $0.007$ in the fixed-$n$ arm), indistinguishable from zero, while the interval width of the \emph{same} credal set reaches $0.519$. The gap between the two summaries, $\drho = +0.54$, is roughly eight times the gap between the credal width and BMA ($+0.067$). This is the location effect hypothesised in \cref{sec:setup}: in the binary case, where the interval sits in the simplex swamps how wide it is.

\emph{The effect weakens as classes are added.} The entropy interval's fidelity rises monotonically from $-0.018$ to $0.500$ between $C=2$ and $C=20$, and its gap to the width falls from $+0.537$ to $+0.348$. This is consistent with its good performance on the many-class image benchmarks on which credal wrappers were introduced \citep{wang2025credal}, where the binary regime was not the focus. At $C=2$ the width equals the credal volume, so our binary result agrees with \citet{sale2023volume}, who find volume a good EU measure for binary problems but weaker with more classes, the opposite of the entropy interval. Credal functionals therefore do not share one regime of validity, and should be chosen for the class count at hand.

\emph{The credal advantage over BMA is also largest at small $C$:} $+0.067$ at $C=2$ ($P<10^{-7}$), not significant by $C=5$, and slightly reversed at $C=20$, in both arms (\cref{app:e1armA}).

\subsection{The credal advantage is a small-ensemble effect}
\label{sec:ensemble}

\Cref{fig:main}b shows the sweep over the number of members (full results in \cref{tab:e2}, \cref{app:perseed}). The credal width leads BMA by $\drho \approx +0.06$ to $+0.07$ at every $M \le 5$ ($P < 10^{-4}$), and wins in 34 to 35 of 40 worlds (\cref{app:perseed}). This is the regime targeted by \citet{wang2025credal}, and it confirms their small-ensemble claim against an oracle reference. The lead is no longer significant at $M=10$ ($P=0.084$) or $M=20$ ($P=0.868$; width wins exactly 20 of 40 worlds), and at $M=40$ BMA is ahead ($\drho = -0.034$, $P=0.010$). The crossover lies between roughly 10 and 20 members. MC-Dropout trails at every $M$, and the entropy interval stays near zero.

\subsection{The advantage does not grow as data thin}
\label{sec:sparsity}

Credal representations are usually motivated by scarce evidence, so one might expect the advantage to grow as the training set shrinks. \Cref{fig:main}c shows otherwise. The width leads BMA significantly at every retained fraction ($P<10^{-3}$ throughout; \cref{app:e3}), but the gap fluctuates between $+0.041$ and $+0.068$ with no trend as $n$ falls from 500 to 100.

If the credal set helped because data are scarce, its advantage would grow as $n$ falls; it does not, yet at fixed $n$ it shrinks steadily as members are added (\cref{sec:ensemble}). The advantage is therefore controlled by $M$, the number of samples describing the model's uncertainty, not by the size of the training set. One untested explanation is that the plug-in mutual-information estimate is noisy at small $M$, while the range is robust at small $M$ but dominated by extreme members at large $M$. We make no claim beyond $n=500$ (\cref{app:wang}).

\subsection{Robustness}
\label{sec:robust}

\paragraph{Do the two estimands agree?} On a fixed design of $n=300$, the disagreement and learning-curve oracles have rank correlation $\rho = 0.761$ across the query grid (\cref{fig:world}, bottom right), so rankings graded against spread largely carry over to reducible error; shared bias concentrates at the edges of dense drilling, where models extrapolate (\cref{fig:world}; \cref{app:oracles}).

\paragraph{Member model.} Repeating the sweep over the number of members with bootstrap MLPs trained on raw coordinates in place of the random-feature members (\cref{app:ablation}) preserves the sign of the small-ensemble gap ($\drho = +0.040$, $P=0.015$ at $M=2$), although the MLP gaps at $M=3$ and $5$ are not individually significant and the MLP run shows no reversal up to $M=40$. The small-$M$ ranking is therefore not an artefact of the random feature map, but the crossover point depends on the member model.

\paragraph{Real geochemistry.} To check that the result is not a property of the simulator, we take the densest $2^\circ$ cell of the South Australian geochemical database \citep{gssa2023sarig} (1{,}218 samples, six lithological classes, nine centred-log-ratio element features), treat it as the population, draw 40 sparse surveys of $n=150$, and compute the oracle from $K=80$ refits on each (\cref{app:realdata}). \Cref{fig:real} (\cref{app:realdata}) shows the same pattern as \cref{sec:ensemble}: the width leads at $M \le 5$ ($P \le 0.002$) and BMA leads from $M=10$. The margin is smaller ($+0.01$ to $+0.03$), as expected at $C=6$ (\cref{sec:functional}).

On the full dataset under spatial block cross-validation, credal and BMA measures of the same backbones are within fold noise in balanced accuracy, and random splits inflate it by about $0.20$ for every method (\cref{app:realdata}).

\section{Discussion and limitations}
\label{sec:discussion}

\paragraph{Practical guidance.} With fewer than about ten members and few classes, read the credal set through its interval width, which is also about four orders of magnitude cheaper than the entropy interval (\cref{app:compute}); avoid the entropy interval in binary problems; with large ensembles or many classes, BMA is at least as good.

\paragraph{Limitations.} (i) The disagreement oracle conditions on the survey design; a design-marginal target would be broader. (ii) The synthetic world is one family of smooth fields, which the real-data oracle only partly mitigates. (iii) The crossover point appears to depend on the member model. (iv) $\Hlo$ is exact only at $C=2$. (v) The credal versus BMA effects are small ($\drho \le 0.07$) though consistent; the functional effect ($\drho \approx 0.5$) is large. (vi) EU is model-specific, so we rank measures, not algorithms. Next, the protocol can compare further credal functionals (volume, non-specificity) against a common oracle.

\section{Conclusions}
\label{sec:conclusion}

We introduced a protocol that grades EU estimators against computed oracles for reducible error and estimator spread, rather than proxies. It applies to any EU estimator whose data-generating process can be simulated. It shows that the credal interval width beats BMA with few ensemble members but not many, that the entropy interval carries no epistemic signal in two-class problems, and that the advantage tracks the number of members, not the amount of data.

\iffalse

\subsection*{AI use statement}
We used generative AI tools to help debug the code used for our experiments and to check the grammar and quality of language in the text. All numerical results were produced by the authors' code, and all AI-assisted code and text were reviewed by the authors. We take responsibility for the final content of this work, including text, claims or artifacts produced with the aid of generative AI.

\subsection*{Ethics statement}
This work uses openly licensed geochemical survey data \citep{gssa2023sarig,usgs2023ngs} and synthetic data; no human subjects or personal data are involved. Better-calibrated exploration models could make drilling more targeted and so reduce ground disturbance, but they may also accelerate extraction whose costs fall on local communities; improved uncertainty estimates do not resolve that asymmetry, and we view such tools as informing rather than replacing expert and community decision-making.

\subsection*{Reproducibility statement}
The generative world, both oracles, all uncertainty measures and the scripts that regenerate every table and figure are provided as anonymised supplementary code. \Cref{alg:protocol} gives the full procedure, all hyperparameters are listed in \cref{app:settings}, the proof of \cref{prop:water} is in \cref{app:proofs}, and real-data preprocessing, including the spatial cross-validation protocol and dataset identifiers, is described in \cref{app:realdata}. Every experiment is seeded, and per-world results are released as CSV files.

\fi

\bibliography{references}
\bibliographystyle{iclr2027_conference}

\appendix
\paragraph{Guide to the appendix.} \Cref{app:proofs} proves \cref{prop:water}; \cref{app:alg} gives the protocol as pseudocode; \cref{app:settings} lists all hyperparameters; \cref{app:e1armA} reports the fixed-$n$ class-count sweep; \cref{app:perseed} gives per-world result distributions; \cref{app:e3} the full sample-size results; \cref{app:wang} a repeat at a larger design; \cref{app:oracles} details of the oracle-agreement experiment; \cref{app:ablation} the member-model ablation; \cref{app:realdata} real-data preprocessing and spatial cross-validation; and \cref{app:compute} computational cost.

\section{Proof of \cref{prop:water}}
\label{app:proofs}

$H(p) = -\sum_k p_k \log_2 p_k$ is separable, strictly concave and symmetric in its coordinates, and $\mathcal{P}$ is the intersection of the box $\prod_k [\pL_k, \pU_k]$ with the hyperplane $\sum_k p_k = 1$, assumed non-empty. The maximiser is therefore unique. The KKT conditions read $-\log_2 p_k - 1/\ln 2 = \lambda + \nu_k - \eta_k$ with $\nu_k, \eta_k \ge 0$ the multipliers for $p_k \le \pU_k$ and $p_k \ge \pL_k$ and complementary slackness. For every coordinate strictly inside its box, $\nu_k = \eta_k = 0$, so all such coordinates share the common value $t = 2^{-\lambda - 1/\ln 2}$. A coordinate at its upper bound has $\nu_k \ge 0$, hence $\pU_k \le t$; a coordinate at its lower bound has $\pL_k \ge t$. In all cases $p_k = \mathrm{median}(\pL_k, t, \pU_k)$. The map $t \mapsto \sum_k \mathrm{median}(\pL_k, t, \pU_k)$ is continuous and non-decreasing, ranging from $\sum_k \pL_k \le 1$ to $\sum_k \pU_k \ge 1$, so a level satisfying the constraint exists and yields the unique maximiser; it is located by bisection (80 iterations). $\square$

\begin{figure}[H]
\centering
\includegraphics[width=0.8\linewidth]{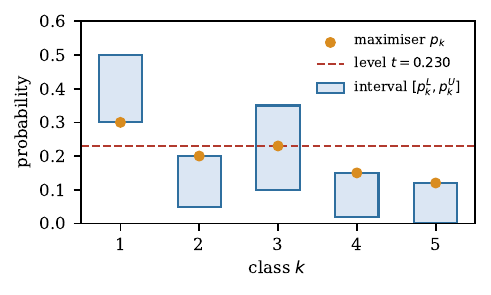}
\caption{Water-filling for the upper entropy with $C=5$. Classes 2, 4 and 5 are held at their upper bounds, class 1 at its lower bound, and class 3 sits at the level $t$.}
\label{fig:waterfill}
\end{figure}

\section{Protocol pseudocode}
\label{app:alg}

\begin{algorithm}[H]
\caption{Oracle-referenced grading of EU measures}
\label{alg:protocol}
\begin{algorithmic}[1]
\Require learning algorithm $\mathcal{A}$; survey size $n$; ensemble size $M$; oracle refits $K$; query grid $G$; uncertainty measures $U_1,\dots,U_J$; number of worlds $S$
\For{$s = 1,\dots,S$}
  \State Draw a world $p^\star$ and a design $X = \{x_1,\dots,x_n\}$; draw labels $y_i \sim \mathrm{Cat}(p^\star(x_i))$
  \Statex \textit{Estimator path}
  \For{$m = 1,\dots,M$}
    \State Resample $(X, y)$ with replacement to get $\mathcal{D}_m$; fit $f_m \gets \mathcal{A}(\mathcal{D}_m)$; set $p_m(x) \gets f_m(x)$ for $x \in G$
  \EndFor
  \State For each measure $j$ and each $x \in G$, compute $U_j(x)$ from $p_1(x),\dots,p_M(x)$
  \Statex \textit{Oracle path}
  \For{$k = 1,\dots,K$}
    \State Draw fresh labels $y^{(k)}_i \sim \mathrm{Cat}(p^\star(x_i))$ at the same design; refit $q_k \gets \mathcal{A}(X, y^{(k)})$
  \EndFor
  \State $\mathrm{EU}^{\mathrm{spr}}(x) \gets H(\bar q(x)) - \frac1K \sum_k H(q_k(x))$ for $x \in G$, where $\bar q = \frac1K\sum_k q_k$
  \State $\rho^{(s)}_j \gets \mathrm{Spearman}\big(U_j, \mathrm{EU}^{\mathrm{spr}}\big)$ over $G$, for each $j$
\EndFor
\State For each pair $(A, B)$ of measures: $\Delta\rho^{(s)} \gets \rho^{(s)}_A - \rho^{(s)}_B$; report the mean, a bootstrap 95\% interval and a Wilcoxon signed-rank $P$-value over $s = 1,\dots,S$
\end{algorithmic}
\end{algorithm}

\section{Settings}
\label{app:settings}

\begin{table}[H]
\centering\small
\caption{Hyperparameters used for all simulation results unless stated otherwise.}
\begin{tabular}{lll}
\toprule
Component & Parameter & Value \\
\midrule
World & domain; bodies per class $B$ & $[0,1]^2$; 3 \\
 & centres; widths; amplitudes & $\mathcal{U}[0.05,0.95]^2$; $\mathcal{U}[0.10,0.28]$; $\mathcal{U}[0.8,2.0]$ \\
 & inverse temperature $\beta$ & 2.5 \\
Survey & hubs; scatter about hub & 4 drawn from $\mathcal{U}[0.12,0.88]^2$; $\mathcal{N}(0, 0.11^2)$, clipped \\
Member model & random features $D$; bandwidth $\gamma$ & 128 (shared across members); 9.0, $W \sim \mathcal{N}(0, 2\gamma)$ \\
 & inverse regularisation; diversity & 1.0 ($\ell_2$); bootstrap of size $n$ \\
Oracle & refits $K$; query grid & 80; $28\times28$ on $[0.02,0.98]^2$ \\
Credal & $\Hup$; $\Hlo$ & bisection, 80 iterations; greedy vertex construction \\
Inference & worlds per cell; test & 40; paired two-sided Wilcoxon signed-rank \\
\bottomrule
\end{tabular}
\end{table}

\section{Class-count sweep: full results and fixed-\texorpdfstring{$n$}{n} arm}
\label{app:e1armA}

\begin{table}[H]
\centering
\small
\caption{Class-count sweep at fixed samples per class ($n = 60C$, $M=5$, 40 worlds). Mean Spearman fidelity to the oracle, and paired gaps with bootstrap 95\% intervals and Wilcoxon $P$-values.}
\label{tab:e1}
\setlength{\tabcolsep}{3.5pt}
\resizebox{\linewidth}{!}{%
\begin{tabular}{rr ccc cc cc}
\toprule
& & \multicolumn{3}{c}{Fidelity $\rho$} & \multicolumn{2}{c}{width $-$ entropy interval} & \multicolumn{2}{c}{width $-$ BMA} \\
\cmidrule(lr){3-5}\cmidrule(lr){6-7}\cmidrule(lr){8-9}
$C$ & $n$ & width & BMA & ent.\ int. & $\drho$ [95\% CI] & $P$ & $\drho$ [95\% CI] & $P$ \\
\midrule
2  & 120  & \textbf{0.519} & 0.452 & $-0.018$ & $+0.537$ [$+0.465$, $+0.610$] & $<10^{-11}$ & $+0.067$ [$+0.049$, $+0.084$] & $<10^{-7}$ \\
3  & 180  & \textbf{0.636} & 0.610 & 0.222 & $+0.414$ [$+0.344$, $+0.481$] & $<10^{-10}$ & $+0.026$ [$+0.012$, $+0.041$] & $<10^{-3}$ \\
5  & 300  & \textbf{0.690} & 0.683 & 0.239 & $+0.451$ [$+0.370$, $+0.530$] & $<10^{-10}$ & $+0.007$ [$-0.008$, $+0.022$] & 0.368 \\
10 & 600  & 0.792 & \textbf{0.797} & 0.360 & $+0.432$ [$+0.377$, $+0.489$] & $<10^{-11}$ & $-0.005$ [$-0.015$, $+0.004$] & 0.397 \\
20 & 1200 & 0.847 & \textbf{0.862} & 0.500 & $+0.348$ [$+0.310$, $+0.385$] & $<10^{-11}$ & $-0.015$ [$-0.026$, $-0.005$] & 0.003 \\
\bottomrule
\end{tabular}}
\end{table}

In this arm $n = 400$ for every $C$, so samples per class fall as $C$ grows (to a minimum class size of 1.3 on average at $C=20$). The entropy interval is again uninformative at $C=2$, and the width $-$ BMA gap again favours the width only at small $C$. The non-monotone recovery of the entropy interval here reflects the collapse of per-class data at large $C$, which is why the fixed-samples-per-class arm is the controlled comparison.

\begin{table}[H]
\centering\small
\caption{Fixed $n=400$, $M=5$, 40 worlds. Mean fidelity (s.d.) and paired gaps.}
\setlength{\tabcolsep}{3.5pt}
\resizebox{\linewidth}{!}{%
\begin{tabular}{r ccc cc cc}
\toprule
$C$ & width & BMA & ent.\ int. & $\drho$ (w $-$ ent) & $P$ & $\drho$ (w $-$ BMA) [95\% CI] & $P$ \\
\midrule
2  & 0.634 (0.167) & 0.568 (0.207) & 0.007 (0.247) & $+0.627$ & $<10^{-11}$ & $+0.067$ [$+0.043$, $+0.093$] & $<10^{-6}$ \\
3  & 0.719 (0.127) & 0.703 (0.131) & 0.255 (0.186) & $+0.464$ & $<10^{-11}$ & $+0.016$ [$+0.001$, $+0.033$] & 0.128 \\
5  & 0.765 (0.103) & 0.758 (0.106) & 0.322 (0.194) & $+0.443$ & $<10^{-11}$ & $+0.007$ [$-0.003$, $+0.017$] & 0.320 \\
10 & 0.750 (0.140) & 0.755 (0.138) & 0.295 (0.191) & $+0.454$ & $<10^{-11}$ & $-0.006$ [$-0.017$, $+0.005$] & 0.320 \\
20 & 0.775 (0.109) & 0.792 (0.116) & 0.233 (0.205) & $+0.542$ & $<10^{-11}$ & $-0.016$ [$-0.029$, $-0.004$] & 0.024 \\
\bottomrule
\end{tabular}}
\end{table}

\section{Number of members: full results and per-world distributions}
\label{app:perseed}

\begin{table}[H]
\centering
\small
\caption{Sweep over the number of members $M$ ($n=150$, $C=2$, 40 worlds): mean fidelity and paired width $-$ BMA gap.}
\label{tab:e2}
\begin{tabular}{r cccc cc}
\toprule
$M$ & width & BMA & ent.\ int. & MC-Dropout & $\drho$ (width $-$ BMA) [95\% CI] & $P$ \\
\midrule
2  & \textbf{0.330} & 0.261 & $-0.000$ & 0.073 & $+0.069$ [$+0.048$, $+0.091$] & $<10^{-6}$ \\
3  & \textbf{0.421} & 0.347 & $-0.036$ & 0.089 & $+0.074$ [$+0.059$, $+0.090$] & $<10^{-9}$ \\
5  & \textbf{0.475} & 0.416 & $-0.052$ & 0.253 & $+0.059$ [$+0.036$, $+0.082$] & $<10^{-4}$ \\
10 & \textbf{0.589} & 0.568 & $-0.024$ & 0.345 & $+0.021$ [$-0.003$, $+0.044$] & 0.084 \\
20 & 0.671 & \textbf{0.675} & 0.054 & 0.380 & $-0.004$ [$-0.026$, $+0.020$] & 0.868 \\
40 & 0.684 & \textbf{0.719} & 0.007 & 0.459 & $-0.034$ [$-0.060$, $-0.011$] & 0.010 \\
\bottomrule
\end{tabular}
\end{table}

\begin{table}[H]
\centering\small
\caption{Per-world distribution of $\drho = \rho_{\mathrm{width}} - \rho_{\mathrm{BMA}}$. Left: sweep over the number of members. Right: class-count sweep, fixed $n$.}
\begin{tabular}{r cccc}
\toprule
$M$ & wins & median & min & max \\
\midrule
2  & 35/40 & $+0.064$ & $-0.073$ & $+0.245$ \\
3  & 35/40 & $+0.080$ & $-0.033$ & $+0.194$ \\
5  & 34/40 & $+0.062$ & $-0.121$ & $+0.302$ \\
10 & 27/40 & $+0.029$ & $-0.160$ & $+0.174$ \\
20 & 20/40 & $+0.001$ & $-0.153$ & $+0.172$ \\
40 & 12/40 & $-0.027$ & $-0.277$ & $+0.088$ \\
\bottomrule
\end{tabular}
\hspace{1.5em}
\begin{tabular}{r cccc}
\toprule
$C$ & wins & median & min & max \\
\midrule
2  & 31/40 & $+0.048$ & $-0.040$ & $+0.304$ \\
3  & 24/40 & $+0.004$ & $-0.071$ & $+0.157$ \\
5  & 19/40 & $-0.001$ & $-0.047$ & $+0.099$ \\
10 & 16/40 & $-0.002$ & $-0.143$ & $+0.076$ \\
20 & 10/40 & $-0.017$ & $-0.115$ & $+0.058$ \\
\bottomrule
\end{tabular}
\end{table}

\section{Sample-size sweep}
\label{app:e3}

\begin{table}[H]
\centering\small
\caption{Retained fraction of a survey of $n=500$ ($M=5$, $C=2$, 40 worlds).}
\begin{tabular}{rr cccc cc}
\toprule
Retained & $n$ & width & BMA & ent.\ int. & MC-Dropout & $\drho$ (width $-$ BMA) [95\% CI] & $P$ \\
\midrule
100\% & 500 & \textbf{0.651} & 0.606 & 0.109 & 0.280 & $+0.046$ [$+0.021$, $+0.070$] & $<10^{-3}$ \\
80\%  & 400 & \textbf{0.622} & 0.570 & 0.089 & 0.308 & $+0.052$ [$+0.034$, $+0.071$] & $<10^{-5}$ \\
60\%  & 300 & \textbf{0.629} & 0.589 & 0.082 & 0.368 & $+0.041$ [$+0.024$, $+0.057$] & $<10^{-4}$ \\
40\%  & 200 & \textbf{0.460} & 0.393 & $-0.054$ & 0.190 & $+0.068$ [$+0.043$, $+0.093$] & $<10^{-5}$ \\
20\%  & 100 & \textbf{0.516} & 0.455 & $-0.045$ & 0.198 & $+0.061$ [$+0.041$, $+0.082$] & $<10^{-6}$ \\
\bottomrule
\end{tabular}
\end{table}

\section{Number of members at two design sizes}
\label{app:wang}

We repeated the sweep over the number of members at $n \in \{150, 600\}$ ($C=2$). At $M=5$, the setting emphasised by \citet{wang2025credal}, the width leads BMA at $n=150$ ($\drho = +0.039$, $P=0.001$); at $n=600$ the gap is not significant ($\drho = -0.009$, $P=0.261$). In both cases BMA overtakes the width between $M \approx 10$ and $M \approx 20$. The $n=600$ result is why \cref{sec:sparsity} restricts its claim to the range $n \in [100, 500]$.

\section{Oracle agreement}
\label{app:oracles}

On a fixed design of $n=300$ drills, the disagreement oracle uses $K=100$ label redraws and the learning-curve oracle uses $K=60$ sub-surveys to estimate the reducible error $\mathrm{RE}(x,n) = \E[(g_n(x) - p^\star(x))^2] = \mathrm{Bias}^2 + \mathrm{Var}$. In this experiment the disagreement oracle is measured by the prediction variance rather than the mutual-information form of \cref{def:spr}. The two oracles have Spearman correlation $0.761$ over the grid. The shared-bias map $\hat B^2 = \max(\mathrm{RE} - \widehat{\mathrm{Var}}, 0)$ is concentrated at the periphery of dense drilling; we report it qualitatively, since a pointwise positive estimate at a given location need not be distinguishable from estimation noise.

\section{Member-model ablation}
\label{app:ablation}

\begin{table}[H]
\centering\small
\caption{Paired gap $\drho = \rho_{\mathrm{width}} - \rho_{\mathrm{BMA}}$ for two member models ($n=150$, $C=2$, 40 worlds). The random-feature model here uses $D=256$ features, lengthscale 0.25; the MLP has 32 and 16 ReLU units, $\ell_2 = 0.05$, on raw coordinates.}
\label{tab:ablation}
\begin{tabular}{r cc cc}
\toprule
& \multicolumn{2}{c}{Random-feature logistic} & \multicolumn{2}{c}{MLP} \\
\cmidrule(lr){2-3}\cmidrule(lr){4-5}
$M$ & $\drho$ & $P$ & $\drho$ & $P$ \\
\midrule
2  & $+0.073$ & $<0.001$ & $+0.040$ & 0.015 \\
3  & $+0.061$ & $<0.001$ & $+0.027$ & 0.090 \\
5  & $+0.028$ & 0.064 & $+0.041$ & 0.070 \\
10 & $-0.013$ & 0.277 & $+0.055$ & 0.030 \\
20 & $-0.026$ & 0.015 & $+0.041$ & 0.044 \\
40 & $-0.043$ & 0.001 & $+0.073$ & 0.004 \\
\bottomrule
\end{tabular}
\end{table}

\section{Real-data preprocessing and deployment results}
\label{app:realdata}

\paragraph{Data.} Drill-hole assays from the South Australian Resources Information Gateway \citep{gssa2023sarig}, released under the South Australian Government open-data licence. Values below detection are imputed at half the detection limit; nine elements (Cu, Si, Al, Fe, Mg, Ca, Na, K, Ti) are retained on coverage; features are centred-log-ratio transformed and standardised on training folds only. The processed data contain 9{,}473 samples in six lithological classes (chance $1/6$).

\begin{table}[H]
\centering\small
\caption{Resampling oracle on real geochemistry (dense SARIG sub-region, $N_{\mathrm{pop}}=1{,}218$, $C=6$, $n=150$, $K=80$, 40 sub-surveys): mean fidelity and paired gap $\drho$ (width $-$ BMA).}
\label{fig:real}
\setlength{\tabcolsep}{5pt}
\begin{tabular}{l cccccc}
\toprule
$M$ & 2 & 3 & 5 & 10 & 20 & 40 \\
\midrule
BMA & 0.409 & 0.528 & 0.653 & \textbf{0.763} & \textbf{0.841} & \textbf{0.902} \\
width & \textbf{0.441} & \textbf{0.555} & \textbf{0.664} & 0.752 & 0.803 & 0.843 \\
$\drho$ & $+0.032$ & $+0.027$ & $+0.011$ & $-0.012$ & $-0.038$ & $-0.059$ \\
$P$ & $<0.001$ & $<0.001$ & 0.002 & $<0.001$ & $<0.001$ & $<0.001$ \\
\bottomrule
\end{tabular}
\end{table}

\paragraph{Resampling oracle.} The densest $2^\circ$ cell ($136$--$138^\circ$E, $34$--$32^\circ$S) serves as the population ($N_{\mathrm{pop}}=1{,}218$). For each of 40 sparse surveys of $n=150$, $K=80$ bootstrap refits give the oracle on the remaining $\approx 1{,}068$ locations, and an RFF-logistic ensemble ($D=256$, lengthscale 1.0) with up to 40 members supplies the measures.

\paragraph{Spatial versus random cross-validation.} Spatial blocks are formed by $k$-means on site coordinates and assigned to five folds \citep{roberts2017cross}.

\begin{table}[H]
\centering\small
\caption{Balanced accuracy under spatial block and random 5-fold CV ($C=6$). Credal rows share their backbone with the BMA row above them; only the measure differs.}
\begin{tabular}{llr ccc}
\toprule
Method & Backbone & Params & Spatial & Random & Leak \\
\midrule
Single NN & MLP 9--128--128--6 & 18{,}566 & 0.398 & 0.620 & $+0.223$ \\
Temp.-scaled NN & MLP $+ T$ & 18{,}567 & 0.398 & 0.620 & $+0.223$ \\
Deep ensemble (BMA) & $5\times$ MLP & 92{,}830 & 0.399 & 0.615 & $+0.216$ \\
Credal (ensemble) & same & 92{,}830 & 0.401 & 0.610 & $+0.209$ \\
MC-Dropout (BMA) & MLP, $p=0.3$ & 18{,}566 & 0.426 & 0.619 & $+0.193$ \\
Credal (MC-Dropout) & same & 18{,}566 & 0.423 & 0.614 & $+0.191$ \\
Variational BNN (BMA) & DenseFlipout & 37{,}132 & 0.431 & 0.596 & $+0.165$ \\
Credal (BNN) & same & 37{,}132 & 0.435 & 0.592 & $+0.157$ \\
\bottomrule
\end{tabular}
\end{table}

\section{Computational cost}
\label{app:compute}

\begin{table}[H]
\centering\small
\caption{Wall-clock cost per world at $n=400$ over 784 query points, single CPU core.}
\begin{tabular}{l ccc}
\toprule
Component & $C=2$ & $C=5$ & $C=20$ \\
\midrule
Oracle, $K=80$ refits & 0.46 s & 0.65 s & 1.33 s \\
Ensemble, $M=5$ & 24 ms & 37 ms & 65 ms \\
Measure: BMA & 0.17 ms & 0.26 ms & 0.50 ms \\
Measure: credal width & 0.03 ms & 0.04 ms & 0.14 ms \\
Measure: entropy interval & 247 ms & 320 ms & 510 ms \\
\bottomrule
\end{tabular}
\end{table}

The completed simulation programme (640 worlds, 51{,}200 oracle refits, about 4{,}000 ensemble fits) runs in under 30 minutes on one CPU core with no accelerator.

\end{document}